\documentclass[12pt, a4paper]{article} 

\usepackage[english]{babel} 
\usepackage{booktabs} 
\usepackage{comment} 
\usepackage[utf8]{inputenc} 
\usepackage[T1]{fontenc} 

\usepackage{amsmath,amsfonts,amsthm} 
\usepackage{mathabx}
\usepackage{txfonts}
\usepackage{stmaryrd}
\theoremstyle{definition}
\newtheorem{definition}{Definition}[section]

\usepackage{tikz} 
\usepackage{tikz-cd}
\usetikzlibrary{positioning,arrows} 
\usetikzlibrary{decorations.pathreplacing} 
\usepackage{tikzsymbols} 
\usepackage{blindtext} 

\usepackage[hang]{footmisc}

\usepackage{geometry} 

\usepackage{setspace} 
\usepackage[T1]{fontenc} 
\usepackage[utf8]{inputenc} 

\usepackage{fancyhdr} 
\usepackage{natbib} 
\usepackage{har2nat} 

\usepackage{hyperref}

\hypersetup{
	colorlinks=true,
	citebordercolor=0 0 0,
}

\newcommand{\N}{\mathbb{N}}

\newcommand\sqin{\mathrel{\raisebox{0.1ex}{\ooalign{$\sqsubset$\cr $-$}}}}
\newcommand\Slv[2][]{\hspace{-0.9pt}{\curlywedgeuparrow^{#1}}{#2}}

\title{Morphological stability and identity in self-organizing systems} 
\author{
	Igor Strozzi 
}
\date{}
\begin{document}
	
	
	\maketitle 
	
	\begin{abstract}
		
		In this essay I aim to investigate and discuss the process through which bundles of things ``self-organize'' into other things. In particular, I engage in such investigation by trying to apply a  framework of analysis of structural stability of morphological forms to the context of a continuous cellular automata known as \textit{Lenia}, which displays great diversity of distinguishable, individuated complex subsystems, who behave autonomously and in lifelike manner (such subsystems are often known as \textit{solitons}). I do so in order to suggest a path to the development of methods to formally distinguish said subsystems from their environment. 
		
	\end{abstract}
	
	\section{Introduction} 
	
	It could be said that this essay aims to deliberate over the following motivating question: what makes a creature in \textit{Lenia} a creature, instead of just another splodge?
	
	I will not delve deeply into what is \textit{Lenia} \cite{Chan2019}, and I will also consider it only in its simplest form. But, to back up the question asked, I will provide a sketch of its definition. For the time being, let us just consider, rather informally, it to be a dynamical system defined by a rectangular region $\mathcal{R}$ of $\mathbb{R}^2$ with toroidal boundary conditions, to which we ascribe, for each $\mathbf{x} \in \mathcal{R}$, a real number $\sigma_{\mathbf{x}}(t) \in [0, 1]$, called the \textit{state} of $\mathbf{x}$ at the instant $t$. The dynamics are given by a local update rule $\phi$, that maps the state $\sigma_{\mathbf{x}}(t)$ of a point $\mathbf{x}$ to its state in the next instant, with dependence on the states found in a neighborhood of $\mathbf{x}$. At each instant, all states are synchronously updated by the application of $\phi$. Both the neighborhoods and the mapping $\phi$ are considered to be constant in time. 
	
	Pretty much all of the above can -- and is, see \cite{Chan2019, Chan2020} -- be generalized, but it suffices (rather, it is more adequate) to the purpose of this work to consider of all cases the simplest. Indeed, the question stated as motivating is too hard already. I do not even faintly hope to answer it, at least not in such a brief essay. My goal, if anything more than briefly digress over millennial issues (such as problems concerning the metaphysics of identity) and other elusive concepts, such as that of self-organization, is to bring a deep and promising work \cite{Thom1982} out of its apparent oblivion, in thrust that it is still actual and can provide useful insight into problems such as the question asked.
	
	\section{Some cautionary observations}
	
	Before properly starting the proposed discussion, I should state that I do not believe that there exists an \textit{ontological} criterion to distinguish between entities, or objects. It thus follows that there is no purely objective way to distinguish between an object and its environment and, as a matter of fact, that the very notion of object is ontologically disputable. This position, evidently, goes against experience and, in fact, is quite inconvenient for probably every practical and most theoretical problems we as humans face. It does not, nonetheless, render this whole discussion pointless, simply because at an epistemological/phenomenological level, individuation is possible.
	
	The relevance of these observations lie in the fact that this position on the fundamental nature of concepts such as identity and individual will necessarily permeate and underlie the development of this work. It should be said that such skepticism, although, as far as I know, unconventional elsewhere, falls not too faraway from the generally accepted idea that the delimitation of a system's -- or, at least, of its description -- boundaries inherently depends on the observer's purpose and convenience. 
	
	A perhaps stronger claim can be made while still sounding reasonable: there is a plurality -- for physical systems, we would guess that uncountably many -- of equivalent descriptions of a same system. It seems to go implicitly with either claims that no such description is identical to the system it aims to describe. It also seems to be the case that the existence and individuation of such a system is presupposed. An observation to be made is that if, in the former claim, we consider the delimitation of the \textit{system} itself -- instead of its description, or a model for it -- to be dependent on the observer, the assumption that it exists and can be individuated is contradictory. That in turn implies that we can only talk about systems in an epistemological/phenomenological level, despite assuming -- or not -- that they exist as \textit{noumena}.
	
	\section{Is self-organization actually a well-defined concept?}
	
	Even to these days, more than 70 years after what seems to be the first appearance of the term ``self-organizing system'' (SOS) \cite{Gershenson2020, Banzhaf2009}, introduced in \cite{Ashby1947}, the concept of self-organization still remains a somewhat elusive idea, lacking a formal \cite{Gershenson2020}, unified definition and, much for this very reason, encompassing a variety of partial, unsystematic and sometimes even contradictory -- in the sense that two different self-organized systems might display mutually exclusive characteristics, e.g., one being taken as deterministic and the other as stochastic, and that can lead to contradictory outlines of the properties inherent to SOSs  -- different characteristics/characterizations.
	
	An interesting question -- among others -- raised in \cite{Collier2004} is: what is this ``self'' that organizes? An answer, seemingly given by Maturana \cite{Collier2004}, is that, as it looks evident, a system cannot organize itself: its selfness arises as it comes to being, through the spontaneous organization of its parts. Another answer, given by Collier at the same work, is that it is reasonable to talk about self-organization (without that being a misnomer) since the relevant factors to the emergence of the system, and thus the ``self'', are internal (to the system?), and not modulated by external influence. I do not feel satisfied with either of these answers, but, to me, the latter is clearly more problematic, although the former seems to imply that ``self-organization'' is in fact an oxymoron, at best -- which I also do not think to be the case.
	
	I do not intend to solve the above dispute or answer if there is a well-defined self that organizes itself, but I will offer a pragmatical and temporary suggestion. A more deeper answer, I believe, can be found investigating in which ways does it make sense to talk about ``selfs'' and ``systems'', as suggested in the last section. As for now, let us just discuss what I consider a more convenient -- to the present discussion -- definition of a system.
	
	\section{The ``selfhood'' of systems}
	
	As mentioned earlier, there are some widely known facts about systems that pose, one could say, intrinsic restrictions in the grail of finding clear, objective methods of individuation. An immediate objection is the fact that every known ``concrete'' system is open, what forces any description to be no more than that. Hence a scientific-realist position, in the context of systems sciences, seems inherently ill-suited.
	
	Of higher importance still is the matter of equivalence of descriptions between ``equal'' systems, which rises as a major hindrance. One ``same'' system can be seem under a variety of levels of description ``same'' system can be seem under a variety of levels of description \cite{Koestler2013, Koestler1978, Haken2006, Mesarovic1970}, which usually lead to distinct composing entities (parts), and different dynamics and relations between them. If selfhood is inherent to such systen, and not some contingent property born from our point-of-view, as observers, and subsequent description, then, in this sense, such selfhood, or, rather, its characterization, surelly cannot rely on pretty much anything internal to the system -- i.e., components; dynamics and relations between them. But then, what else remains to provide such characterization?
	
	In various references \footnote{Most of our references discuss these issues. For more comprehensive views see \citet{Fieguth2017, Mobus2015, Takahashi2010, Haken2006}} the above problems are considered and one could, as it seems, reduce them under two kinds of equivalence categories -- I am intentionally not using the more technical term ``classes'' -- which I will call \textit{vertical} and \textit{horizontal}. The former category is related to the equivalence between levels of description ``same'' system can be seem under a variety of levels of description \cite{Haken2006, Mesarovic1970}. A system of molecules is also a system of atoms. The latter relate sto the fact that, at least for real/concrete systems, what separates a system from its environment is oftentimes unclear. What seems to be the case -- and I risk saying that this is \textit{always} the case -- is that systems are nested within other systems. Said system of molecules ($M$) is also a system of atoms $A$, each of which is subsystem $A_i$ relatively to $M$. Within $M$, there is an environment wherein the $A_i$ interact, albeit simple such environment can possibly be. This is nothing new, and various works concern such matters (cf. \citet{Walloth2016, Koestler2013}). Yet, in a work still under development -- from which I will borrow some ideas -- I consider such matters under the concept of holons and holarchies, aiming to properly define these two modes of equivalence and, what seems to me even more elusive, the way they relate to each other.
	
	Indeed, it seems an ubiquitous fact that vertical and horizontal equivalences are deeply intertwined. Considering a ``spatial resolution'' parameter $\alpha$, through which variation one obtains a continuum of levels of description of a ``same'' system, the horizontal equivalence problem naturally arises. 
	
	For an example, consider a metallic sphere $S$. At a macroscopic level -- order of magnitude of $10^\alpha, \ \alpha = 1$ meter --, one could reasonably say that the ball is clearly individuated from its environment, having a well-defined surface which separes it from its surroundings. Were one vary $\alpha$ to the characteristic length of electron interactions -- roughly $\alpha = -15$, one would find a rather distinct picture. The surface that once isolated the $S$ so conspicuously, now (heavily?) interacts with the environment. 
	
	We could, naively, say that at the very least some electrons leave the sphere, whereas others become part of it. Even though effects resulting from these dynamics might be negligible at macroscale, there are other interactions at such microscales which are not.
	
	In any case, such electrons are themselves substructures/subsystems of $S$, at least until... they are no more. Conversely, free electrons in the environment are not part of the system until captured by it. One could imagine, thus, that there is, in fact, a diffuse boundary, a gradient, governing the (in-out)flux of electrons through $S$'s ``surface'', since it is not entirely clear or determinable when an electron becomes or stops being part of $S$. To sum it up, how ``closed'' a system appears to be seems, frequently, to depend on the level of description one is using.
	
	\section{Some suggestions as to how define a system}
	
	We will proceed very directly to the definitions, and try to explain them as succintly as possible. An auxiliary concept:
	
	\begin{definition}[Valued relation] Let $S = \{S_i\}_{i\in I}$ be a family of sets and $R$ a $n$-ary relation over $S$, that is, $R \subseteq \prod_{i \in I} S_i$. Let $V$ be another set, said the set of values. If $\mathrm{v}$ is a function such that $\mathrm{v}: R \rightarrow V$, we say that $\mathrm{v}(R) = \{\mathrm{v}(r) \in V\ |\ r \in R\}$, which we shall also denote by $(R, V, \mathrm{v})$, is an $n$-ary $\mathrm{v}$-valued relation over $S$, or a $\mathrm{v}$-valuation of $R$, and, naturally, say that $\mathrm{v}$ is a valuation of $R$. If $R_\mathrm{v} = (R, V, \mathrm{v})$, we say that $R$ is the underlying relation of $R_\mathrm{v}$. $\Diamondblack$
	\end{definition}
	\noindent
	
	In the above definition, we consider the graph of $\mathrm{v}$ as the valued relation, which is very natural since, well, graphs of functions are indeed relations. Thus a $n$-ary valued relation over a family $\{S_i\}$ is nothing else than a $(n+1)$-ary relation with some particularities: mainly, that it is functional in $V$, but it will be almost always the case here that the set $V$ is very different in nature from the $S_i$. 
	
	We will usually talk about relations defined over the elements of a system, which will be considered as sets themselves, and the values these relations have will usually be numbers. The underlying relations account for the `structure of interactions', whereas the valuations denote the characteristics these interactions have, the most remarkable one probably being their strength, or intensity. `Structure of interactions with values' could also be thought of, very naturally, as a graph with weighted edges.
	
	As for the concept of system, there are some comments to be made before we provide our definition. The usual ``minimal'' \cite{Mesarovic1989} set-theoretical definition considers an abstract system to be a relation $S \subseteq \prod_i V_i$, where the $V_i$ represent the objects which are parts of $S$, in a sense that were one to talk about `elements' of a system, these would not necessarily be the $V_i$. We shall similarly conceive a system in a set-theoretical manner, but starting with a single set, which we will call the \textit{underlying set} of the system. This set, though, is not to be understood as the set of `objects', or `elements', of the system. Rather, such elements will be disjoint subsets of the underlying set. 
	
	The above choices are in consonance with our inclination towards spatial systems and the idea of ``leaving some space'' for the environment containing the actual system: under such interpretation, the underlying set would correspond to a spatial region, and the system's elements would correspond to subregions with ``properties'' -- the system itself being the whole of these elements, their properties and relations between them, not including the region itself, although fundamentally depending on it. 
	
	We should further add that despite the possibility of underlying sets having elements of arbitrary nature, the systems in which we are interested are those that arise from the things that are built upon the underlying set, with the nature of its elements being in general of little importance. This choice is to emphasize, mainly, the mereological aspects of our understanding of systems (and the topological ones, which in fact are going to be the most relevant in the present work). The following definition is also intented as to provide suitable ways of allowing one to use tools that benefits from defining hierarchies based on set-inclusion, rather than set-membership (such as measure-theoretical and topological tools). 
	
	\begin{definition}[System] \label{sys} Let $S$ be a set with cardinality $|S| \geq 2$. Consider a partition $\mathfrak{P}$ of $S$, and a subset $\Gamma_{\mathfrak{P}}$ of $\mathfrak{P}$, with $|\Gamma_{\mathfrak{P}}| = C$. Define $\mathbf{R} = \{R_k \ | \ k \in \N_{\leq C} \}$ where each $R_k$ is a set of $n_k$ $k$-ary relations over $\Gamma_{\mathfrak{P}}$, i.e., for any $R_{ki} \in R_k$ we have $R_{ki} \subseteq \Gamma_{\mathfrak{P}}^k$ and therefore $R_k = \{R_{ki} \subseteq \Gamma_{\mathfrak{P}}^k\ | \ i \in \N_{\leq n_k}\}$.  Define, now, for each $R_k$ and each relation $R_{ki} \in R_k$, a set $P_{ki}$ (possibly empty) of functions $\mathrm{v}_{kij}$ such that, for each $j$, $\mathrm{v}_{kij}: R_{ki} \rightarrow V_{kij}$, where $V_{kij}$ is the set of values for the valued relation $\mathrm{v}_{kij}(R_{ki})$. Finally, define the set $\mathbf{V} = \{P_{ki} \ | \ k \in \N_{\leq C};\ i \in \N_{\leq n_k} \}$ of sets of valuations. Then a system $\mathcal{S}$ is a quadruple
		\[ \mathcal{S} = (S,\ \Gamma_{\mathfrak{P}}, \ \mathbf{R},\ \mathbf{V}).\]
		
		\noindent
		We call the set $S$ the underlying set of $\mathcal{S}$, denoted by $\mathfrak{u}(\mathcal{S})$. We call $\Gamma_{\mathfrak{P}}$ the set of elements of $\mathcal{S}$, which we denote by $\mathfrak{e}(\mathcal{S})$. We call $\mathbf{R}$ the set of relations of $\mathcal{S}$, denoted $\mathfrak{r}(\mathcal{S})$, and $\mathbf{V}$ the set of valuations of $\mathcal{S}$, denoted $\mathfrak{v}(\mathcal{S})$. Strictly speaking, $\mathbf{V}$ is, as said, a set of sets of valuations, since for each relation in $\mathbf{R}$ we might have multiple distinct valuations; with respect to a system $\mathcal{S}$, though, we simply call it its set of valuations. 
		
		We say that a non-empty set $s \subset S$ is an element of a system $\mathcal{S}$ iff $s \in \Gamma_{\mathfrak{P}}$, and denote the system-membership relation by $\sqin$. Naturally, it is thus true that: $s \sqin \mathcal{S} \iff s \in \Gamma_{\mathfrak{P}}$. $\Diamondblack$ 
	\end{definition}
	
	Here, I will consider the particular case where $S = (X, \mathcal{T})$ is a topological space. With this, we alter definition $5.2$ so that $\Gamma_{\mathfrak{P}} \subset \mathcal{T}; \ \bigcap \Gamma_{\mathfrak{P}} = \emptyset$.
	
	We need two more definitions. 
	
	\begin{definition}[Subsystem] \label{subsys}
		Let $\mathcal{S} = (S, \Gamma_{\mathfrak{P}}, \mathbf{R}, \mathbf{V})$ be a system. Consider the set $\mathfrak{R} = (\bigcup_k \mathbf{R})\cup (\bigcup_i \bigcup_k \mathbf{V})$. A subsystem $\mathcal{S}' = (S',\Gamma_{\mathfrak{P'}}', \mathbf{R'}, \mathbf{V'})$ of $\mathcal{S}$ is a system such that $S' \subseteq S$, $ \bigcup\Gamma_{\mathfrak{P'}}' \subseteq \bigcup\Gamma_{\mathfrak{P}}$, 
		and for which there are families of functions $\{f_n\}, \{g_m\},\ n, m \leq |\Gamma_{\mathfrak{P'}}'| = C'$, such that, for some $p_n, q_n$,  $f_n: \mathfrak{R}^{p_n} \rightarrow \Gamma_{\mathfrak{P'}}'^{q_n}$ and for $p_m, q_m$, $g_m: \mathfrak{R}^{p_m} \rightarrow \mathcal{P}(\Gamma_{\mathfrak{P'}}'^{q_m} \times V_{kij}')$, for given sets of values $V_{kij}'$, satisfying:
		
		i) for any element $R'$ of $\bigcup_k \mathbf{R}'$ there is exactly one $n$ such that for some $\boldsymbol{\sigma} \in \mathfrak{R}^{p_n}$, $f_n(\boldsymbol{\sigma}) = R'$;
		
		ii) for any element $R'$ of $\bigcup_k \mathbf{R}'$, if there is a valuation $\mathrm{v}_{kij}'$ whose domain $\mathrm{dom}(\mathrm{v}_{kij}')$ is $R'$, then there is exactly one $m$ such that for some $\boldsymbol{\sigma} \in \mathfrak{R}^{p_m}$, $g_m(\boldsymbol{\sigma}) = \mathrm{v}_{kij}'$. $\Diamondblack$
	\end{definition}
	
	In good english, what the definition above means is that a subsystem $\mathcal{S}' \subset \mathcal{S}$ has as its underlying set $S'$ a subset of the underlying set $S = \mathfrak{u}(\mathcal{S})$, and that its relations and valuations depend on the set $\mathfrak{R}$ of all relations/valuations of $\mathcal{S}$. It is too restrictive to aprioristically assume basically anything about how relations on a subsystem relate to relations on its supersystem. In principle, a relation on a subsystem could depend in arbitrarily complicated ways on relations of its supersystem. We cannot even guarantee that a relation $R$ of $\mathcal{S}'$ that is mapped from $\boldsymbol{\sigma} \in \mathfrak{R}$ and has a valuation $\mathrm{v}$ satisfy something like $g_m(\boldsymbol{\sigma}) = \mathrm{v}$, i.e., if $R$ comes from some $\boldsymbol{\sigma}$, its valuations not necessarily depend on the same $\boldsymbol{\sigma}$.
	
	Indeed, an unvalued relation represents a `structure of interactions' and as such could be the same for very different valued relations. For instance, two global coupling parameters can describe very distinct features of a system, and have very different values, but would share a same unvalued (unary) relation. It is not unreasonable to consider, then, that, being these features sufficiently different in quality, their dependence on relations at the superlevel could be different enough as to render, at the very least, inconvenient to consider them functions of a same tuple $\boldsymbol{\sigma}$. 
	
	Let us now define a supersystem.
	
	\newcommand{\ssys}{\subset_{sys}}
	
	A really useful definition of supersystem does not follow all that trivially from definition \ref{subsys}. Conceptually, the reason is that while one can use the knowledge about a system to infer a characterization of a subsystem, when only the latter is known it is in general much harder to completely -- or, at least, satisfactorily -- describe some of its supersystems. The more heterogeneous the system, the harder. 
	
	More formally, the motive is that the functions that define a subsystem from a given system do not necessarily have inverses, so that when we define a subsystem, we not necessarily define how to obtain a corresponding supersystem (which is not even unique). It is, though, obvious to think, for a given system $\mathcal{S}'$, of a supersystem of it as a system $\mathcal{S}$ such that $\mathcal{S}' \sqsubset \mathcal{S}$. Moreover, it is reasonable, and more adequate to our purposes, to define a supersystem for sets of systems, despite the fact that such sets could be singletons. We define supersystems as follows:
	
	\begin{definition}[Supersystem] \label{def::Ssys}
		Let $\mathfrak{S} = \{\mathcal{S}_i\}_{i\in I}$ be a family of systems. Define, for each $\mathcal{S}_i$, the set $\mathfrak{R}_i = (\bigcup \mathfrak{r}(\mathcal{S}_i)) \cup (\bigcup \bigcup \mathfrak{v} (\mathcal{S}_i))$. Define $\mathfrak{R} = \bigcup_{i\in I} \{\mathfrak{R}_i\}$. A system $\mathcal{S}$ is said a supersystem of each $\mathcal{S}_i$ (and, by extension, of $\mathfrak{S}$) if it satisfies the following:
		
		i) $\bigcup\limits_{s \in \mathfrak{S}} \mathfrak{u}(s) \subseteq \mathfrak{u}(\mathcal{S})$;
		
		ii) $\bigcup\limits_{s \in \mathfrak{S}} \mathfrak{e}(s) \subseteq \mathfrak{e}(\mathcal{S})$; 
		
		iii) for each $R \in \bigcup\mathfrak{r}(\mathcal{S})$, there is a natural number $p \leq |\mathfrak{R}|$ and a function $f$ such that, for some $\boldsymbol{\sigma} \in \mathfrak{R}^p$, $R = f(\boldsymbol{\sigma})$;
		
		iv) for each $V \in \bigcup\bigcup\mathfrak{v}(\mathcal{S})$, there is a natural number $q \leq |\mathfrak{R}|$ and a function $g$ such that, for some $\boldsymbol{\sigma} \in \mathfrak{R}^q$, $V = g(\boldsymbol{\sigma})$. $\Diamondblack$
	\end{definition}
	
	Lastly, we define a superlevel of a system. 
	
	\begin{definition}[Superlevel (of a) system] \label{def::Slevel}
		Let $\mathcal{S}$ be a system. We say that another system, $\mathcal{S}^+$, is on the (immediate) superlevel of $\mathcal{S}$ if there is a family $\{S_i\}$ of subsystems of $\mathcal{S}$ such that $\mathcal{S} \sqsubset \mathcal{S}^+$, i.e., $\mathcal{S}^+$ satisfies the conditions given at definition \ref{def::Ssys}. We denote the immediate superlevel of a system $\mathcal{S}$ by $\Slv{\mathcal{S}}$ or, sometimes, by $\mathcal{S}^+$. $\Diamondblack$
	\end{definition}
	
	All of that put, we can finally say that self-organization produces individuation through the supervenience of a supersystem, in a superlevel, over its composing subsystems, which are, themselves, selves. Thus we answer the problem of individuation by stating that it tends to, or perhaps necessarily do so, incur in the production of an infinitely descending chain of ``subselves''. 
	
	\section{Forms and structural stability}
	
	In this section, I will define, exactly as in \cite{Thom1982}, the ideas of \textit{form} and \textit{structural stability}.
	
	\begin{definition}[Form]
		Let $E$ be a topological space and $G$ a group (or pseudogroup) acting on $E$. A $G$-form is defined as an equivalence class of closed sets of $E$ modulo the action of $G$.
	\end{definition}
	
	To put the above into context, consider $E$ to be $\mathbf{S} = \mathbb{R}^2 \times [0, 1]$ with the usual euclidean norm (which we will denote by $\| . \|$). Consider $\tau$ to denote the set of open sets of $\mathbb{R}^2$. Let $G$ be a pseudogroup with homeomorphisms $\Gamma = \{ \gamma : s_1 \mapsto s_2 \ | \ s_1, s_2 \in \tau \}$. Returning to \textit{Lenia}, we have that, for each configuration $A^t : \mathbb{R}^2 \mapsto [0, 1]$ we can consider a point in $E$ (cf. \citet{Chan2019}). Such a point corresponds to a configuration. I believe that Thom stated the definition considering $G$ to be a group -- instead of a pseudogroup --, since otherwise the equivalence class appears to be ill-defined (what would be ``an equivalence class of closed sets of $E$ modulo the action of $G$'' if $G$ were a pseudogroup?).
	
	This observation notwithstanding, let us now try to understand what intuitively this definition means. Considering that $G$ is a pseudogroup and defines a quotient space of $E$ (the orbit space), a $G$-form can be seen as given by the equivalence relation $\sim_G$ that identifies open sets in the configuration space. What one has, roughly speaking, is that open sets of different, but neighboring, configurations are mapped onto the same orbits through the configuration space. That gives a way to talk about ``fuzzy boundaries'' for regions/subsystems in respect to their evolution in time.

	\begin{definition}[Structural stability]
		A $G$-form $A$ is said \textit{structurally stable} if any form $B$ sufficiently close to $A$ in $E$ is $G$-equivalent to $A$. Equivalently, a $G$-equivalence class $F$ defines a structurally stable form if and only if the set of points in $E$ pertaining to $F$ is open.
	\end{definition}
	
	The above definition looks a bit confusing, since a form $B$, without respect to any group or pseudogroup, is hitherto undefined. I will consider that $B$ is defined with respect to any other suitable (pseudo)group, which seems the most obvious thing to do. In this sense, we can talk about arbitrary forms and define a specific $G$-form $A$, in the case we are considering, to be structurally stable if there is some $\varepsilon_0 \in \mathbb{R}_{> 0}$ such that for any other $\varepsilon \leq \varepsilon_0$ it holds true that if
	\[\sup_{c \in A \cup B} | \inf_{a \in A} \|c - a \| - \inf_{b \in B} \|c - b \| | < \varepsilon
	\]
	
	then $a \sim_G b, \ a \in A,\ b \in B$.
	
	\section{Morphological stability as an identity criterion?}
	
	Finally, let us consider the following:
	
	\begin{enumerate}
		\item Creatures in \textit{Lenia} correspond to regions in $\mathbb{R}^2$, the regions they occupy;
		\item These regions can be divided infinitely and arbitrarily. One can reasonably assume that, in some random initialization that yields a creature, at some instant, it is possible to define a region $\mathcal{R} \subset \mathbb{R}^2$ that will contain the creature and some parts of such region that will, by the very nature of the rules that define \textit{Lenia}, interact in an autopoietic fashion, i.e., so as to dynamically sustain a set of relations invariant in time \cite{Maturana1980};
		\item Regions as such correspond to open sets of a point in the configuration space $\mathbb{R}^2 \times [0, 1]$;
		\item For a suitable choice of a pseudogroup $G$, the configuration space is partitioned precisely in the orbit space that makes said regions/creatures/systems structurally stable, hence morphologically stable;
		\item It thus individuates such creature, despite allowing fuzzy boundaries, since we can disturb it without breaking its stability.
	\end{enumerate}
	
	\section{Conclusions}
	
	To wrap it up, despite the inherent difficulties in defining boundaries for objects, entities or systems, and the stated skepticism in the possibility of an ontological criterion of discernibility, we aimed to show some tools given at \citet{Thom1982} that might help is in such distinctions. By their very nature -- i.e., since the choice of the (pseudo)group $G$ is arbitrary --, they actually correspond to our view that distinctions of such nature are inherently epistemological.
	
	The idea of self-organization was briefly discussed and, again, problems, that I believe related to the elusive nature of the concept of identity, concerning the discernibility of entities. A simple modification on the definition of system given by \citet{Mesarovic1989} was proposed, with intent of suggesting a path out of the issue posed.
	
	The concepts of form and structural stability, as found in \citet{Thom1982}, were discussed and an attempt to apply them to \textit{Lenia}'s case was made. The relations between all previous sections were outlined, at last.

	\newpage 
	
	\pagenumbering{roman} 
	
	\bibliography{lit.bib} 

@article{Chan2019,
archivePrefix = {arXiv},
arxivId = {1812.05433},
author = {Chan, Bert Wang-Chak},
doi = {10.25088/ComplexSystems.28.3.251},
eprint = {1812.05433},
issn = {08912513},
journal = {Complex Systems},
number = {3},
pages = {251--256},
title = {{Lenia: Biology of artificial life}},
volume = {28},
year = {2019}
}

@book{Fieguth2017,
address = {Cham},
author = {Fieguth, Paul},
booktitle = {An Introduction to Complex Systems},
doi = {10.1007/978-3-319-44606-6},
isbn = {978-3-319-44605-9},
publisher = {Springer International Publishing},
title = {{An Introduction to Complex Systems}},
url = {http://link.springer.com/10.1007/978-3-319-44606-6},
year = {2017}
}

@article{Chan2020,
archivePrefix = {arXiv},
arxivId = {2005.03742},
author = {Chan, Bert Wang-Chak},
doi = {10.1162/isal_a_00297},
eprint = {2005.03742},
journal = {arXiv},
month = {may},
title = {{Lenia and Expanded Universe}},
url = {http://arxiv.org/abs/2005.03742},
year = {2020}
}

@book{Walloth2016,
address = {Cham},
author = {Walloth, Christian},
booktitle = {Emergent Nested Systems},
doi = {10.1007/978-3-319-27550-5},
isbn = {978-3-319-27548-2},
issn = {1860-0832},
publisher = {Springer International Publishing},
series = {Understanding Complex Systems},
title = {{Emergent Nested Systems}},
url = {link.springer.com/10.1007/978-3-319-27550-5},
year = {2016}
}

@book{Haken2006,
author = {Haken, Hermann},
doi = {10.1007/3-540-33023-2},
edition = {3},
isbn = {978-3-540-33021-9},
publisher = {Springer-Verlag Berlin Heidelberg},
series = {Springer Series in Synergetics},
title = {{Information and Self-Organization}},
url = {http://link.springer.com/10.1007/3-540-33023-2},
year = {2006}
}

@article{Takahashi2010,
author = {Takahashi, Shingo and Takahara, Yasuhiko},
isbn = {9781849965125},
issn = {01708643},
journal = {Lecture Notes in Control and Information Sciences},
title = {{Logical Approach to Systems Theory}},
volume = {404},
year = {2010}
}

@book{Mesarovic1970,
address = {New York, NY},
author = {Mesarovic, M.D.},
editor = {Mesarovic, M. D. and Macko, D. and Takahara, Y.},
isbn = {9780124915503},
publisher = {Academic Press Inc.},
title = {{Theory of Hierarchical, Multilevel, Systems}},
year = {1970}
}

@book{Mobus2015,
address = {New York, NY},
author = {Mobus, George E and Kalton, Michael C},
doi = {10.1007/978-1-4939-1920-8},
isbn = {978-1-4939-1919-2},
pages = {782},
publisher = {Springer New York},
series = {Understanding Complex Systems},
title = {{Principles of Systems Science}},
url = {http://www.springer.com/series/5394 http://link.springer.com/10.1007/978-1-4939-1920-8},
year = {2015}
}

@book{Koestler1978,
	author = {Arthur Koestler},
	title = {Janus: A Summing Up},
	year = {1978},
	publisher = {Vintage Books}
}

@article{Koestler2013,
author = {Koestler, Arthur},
doi = {10.4324/9781315014272},
isbn = {9781136446290},
journal = {The Rules of the Game: Interdisciplinarity, Transdisciplinarity and Analytical Models in Scholarly Thought},
number = {2},
pages = {233--248},
title = {{Beyond atomism and holism -- the concept of the holon}},
volume = {13},
year = {2013}
}

@book{Mesarovic1989,
address = {Berlin/Heidelberg},
author = {Mesarovic, M.D. D and Takahara, Y.},
booktitle = {Lecture Notes in Control and Information Sciences},
doi = {10.1007/BFb0042462},
editor = {Mesarovic, Mihajlo D. and Takahara, Yasuhiko},
isbn = {3-540-50529-6},
issn = {01708643},
publisher = {Springer-Verlag},
series = {Lecture Notes in Control and Information Sciences},
title = {{Abstract Systems Theory}},
url = {http://link.springer.com/10.1007/BFb0042462},
volume = {116},
year = {1989}
}

@incollection{Banzhaf2009,
address = {New York, NY},
author = {Banzhaf, Wolfgang},
booktitle = {Encyclopedia of Complexity and Systems Science},
doi = {10.1007/978-0-387-30440-3_475},
issn = {00213292},
number = {2},
pages = {8040--8050},
publisher = {Springer New York},
title = {{Self-organizing Systems}},
url = {http://link.springer.com/10.1007/978-0-387-30440-3{\_}475},
volume = {2},
year = {2009}
}

@article{Gershenson2020,
archivePrefix = {arXiv},
arxivId = {1903.07456},
author = {Gershenson, Carlos and Trianni, Vito and Werfel, Justin and Sayama, Hiroki},
doi = {10.1162/artl_a_00324},
eprint = {1903.07456},
issn = {15309185},
journal = {Artificial Life},
number = {3},
pages = {391--408},
pmid = {32697161},
title = {{Self-organization and artificial life}},
volume = {26},
year = {2020}
}

@article{Ashby1947,
author = {Ashby, W. R.},
doi = {10.1080/00221309.1947.9918144},
issn = {19400888},
journal = {Journal of General Psychology},
number = {2},
pages = {125--128},
pmid = {20270223},
title = {{Principles of the self-organizing dynamic system}},
volume = {37},
year = {1947}
}

@book{Maturana1980,
address = {Dordrecht},
author = {Maturana, Humberto R. and Varela, Francisco J.},
booktitle = {Philosophical Studies},
doi = {10.1007/978-94-009-8947-4},
isbn = {978-90-277-1016-1},
issn = {0031-8116},
pages = {248--254},
publisher = {Springer Netherlands},
series = {Boston Studies in the Philosophy and History of Science},
title = {{Autopoiesis and Cognition}},
url = {http://link.springer.com/10.1007/978-94-009-8947-4},
volume = {42},
year = {1980}
}

@book{Thom1982,
address = {United States},
  title={Structural Stability And Morphogenesis},
  author={Thom, R.},
  isbn={9780429961571},
  year={2018},
  publisher={CRC Press},
  url={https://www.perlego.com/book/1597022/structural-stability-and-morphogenesis-pdf}
}

@article{Collier2004,
author = {Collier, John},
year = {2004},
month = {04},
pages = {},
title = {Self-organization, Individuation and Identity},
volume = {228},
isbn = {9071868788},
journal = {Revue internationale de philosophie}
}
	\bibliographystyle{apsr} 
	
	
\end{document}